\documentclass[twocolumn]{aastex62}
\shorttitle{}
\shortauthors{Zhang}

\begin{document}


\title{\bf A Quantitative Assessment of Communicating Extra-Terrestrial Intelligent Civilizations in the Galaxy and the Case of FRB-like Signals}


\author{Bing Zhang}
\affil{Department of Physics and Astronomy, University of Nevada Las Vegas, NV 89154, USA}
\email{zhang@physics.unlv.edu}



\begin{abstract}
A formula is proposed to quantitatively estimate the signal emission rate of Communicating Extra-Terrestrial Intelligent civilizations (CETIs) in the Galaxy. I suggest that one possible type of CETI signal would be brief radio bursts similar to fast radio bursts (FRBs). A dedicated search for FRB-like artificial signals in the Galaxy for decades may pose a meaningful upper limit on the emission rate of these signals by CETIs. The Fermi-Hart paradox is answered in terms of not having enough observing times for this and other types of signals. Whether humans should send FRB-like signals in the far future is briefly discussed.
\end{abstract}





\section{Drake equation and its manipulation}

In 1961, Frank Drake wrote the famous equation to estimate the number of actively Communicating Extra-Terrestrial Intelligent civilizations (CETIs) in the Milky Way galaxy. The equation reads \citep[see, e.g.][]{vakoch15}
\begin{equation}
 N = R_* \cdot f_p \cdot n_e \cdot f_l \cdot f_i \cdot f_c \cdot L.
 \label{eq:Drake}
\end{equation}
The meaning of each parameter, as defined in the book ``The Drake Equation'' \citep{vakoch15}, reads
\begin{itemize}
 \item $R_*$: Rate of formation of stars suitable for the development of intelligent life;
 \item $f_p$: Fraction of stars with planetary systems;
 \item $n_e$: Number of planets, per solar system, with an environment suitable for life;
 \item $f_l$: Fraction of suitable planets on which life actually appears;
 \item $f_i$: Fraction of life-bearing planets on which intelligent life emerges;
 \item $f_c$: Fraction of civilizations that develop a technology that releases detectable signs of their existence into space;
 \item $L$: Length of time such civilizations release detectable signals into space.
\end{itemize}
In this equation, the first three parameters, $R_*$, $f_p$, and $n_e$, can be constrained from astronomical observations, but the remaining four parameters are related to biology, sociology, and even philosophy, which cannot be quantitatively assessed based on observations of the sole intelligent life as we know it: our own \citep[e.g.][]{burchell06,vakoch15}. It is worth emphasizing that $f_l$, $f_i$, and $f_c$ as appeared in the Drake equation should mean the fractions that life, intelligent life and communicative intelligent life that {\em can eventually} appear (even if in the far future), not the fractions already appeared {\em at the current age of the universe} (which will be discussed as $f'_l$, $f'_i$, and $f'_c$ later). To avoid ambiguity, I believe that it would be better to define them as
\begin{itemize}
 \item $f_l$: Probability of suitable planets that eventually develop life;
 \item $f_i$: Probability of life-bearing planets that eventually develop intelligent life;
 \item $f_c$: Probability that intelligent life that eventually develop technology and become CETIs.
\end{itemize}
In the following, I  define
\begin{equation}
 n_e^{\rm ceti} = n_e f_l f_i f_c,
 \label{eq:neceti}
\end{equation}
which has the physical meaning
\begin{itemize}
 \item $n_e^{\rm ceti}$: Number of planets, per solar system, with an environment suitable for life and eventually develop CETIs.
\end{itemize}
According to the so-called ``astrobiological Copernican principle'' assumption \citep[e.g.][]{westby20}, an Earth-like planet that are suitable for developing life will eventually develop life, intelligent life, and CETIs, so that $f_l f_i f_c$ should be close to unity, i.e. $n_e^{\rm ceti} \lesssim n_e$.

The star formation rate $R_*$ in Eq.(\ref{eq:Drake}) was introduced to describe a steady-state process. Logically it is not straightforward to see how it enters the problem and how $L$ is used to cancel out the ``per unit time'' dimension introduced in $R_*$. Since one is interested in the CETIs who are sending off signals to Earth ``now'' (corrected for the light travel time from the source to Earth), it is more reasonable to start with the {\em number of stars}  in the galaxy now, i.e. $N_*$, rather than $R_*$ to estimate $N$, so that
\begin{equation}
 N = N_* \cdot f_p \cdot n_e^{\rm ceti} \cdot f'_l \cdot f'_i \cdot f'_c,
 \label{eq:Drake2}
\end{equation}
where the three new $f$ parameters are defined as
\begin{itemize}
 \item $f'_l$: Fraction of planets that can in principle develop CETIs and have actually developed life now;
 \item $f'_i$: Fraction of the above-defined life-bearing planets that have developed intelligent life now;
 \item $f'_c$: Fraction of the above-defined intelligent life that have become CETI now.
\end{itemize}
Notice that they are different from $f_l$, $f_i$, and $f_c$ that enter the Drake equation (\ref{eq:Drake}). Let us further define the following four time scales: 
\begin{itemize}
\item $L_p$: the average lifetime of  Earth-like planets (from its birth to death, likely associated with the death of the host star). For Earth, it is at least $\sim 4.54$ billion years; 
\item $L_l$: the average lifetime of life (from its birth to death - probably associated with the death of the planet). For Earth, it is at least $\sim 3.8$ billion years; 
\item $L_i$: the average lifetime of the intelligent life (from its birth to death - could be associated with the death of the planet, but could be sooner due to its self-destroy). For Earth, it is at least $\sim 10^4$ years; 
\item $L_c$: the average lifetime of CETIs (from its birth to death - again could be associated with the death of the planet or its self-destroy). For Earth, it is at least $\sim 10^2$ years, i.e. since humans have developed the technology to send off artificial signals to space. 
\end{itemize}
The term ``average'' in the above definitions refers to the {\em geometric mean} rather than the arithmetic mean, or the average in the logarithmic space. This way, one can write the three fraction parameters defined above as
\begin{eqnarray}
    f'_l & = & \frac{L_l}{L_p}, \label{eq:fl} \\
    f'_i & = & \frac{L_i}{L_l}, \label{eq:fi} \\
    f'_c & = & \frac{L_c}{L_i},
\label{eq:fc}
\end{eqnarray}
so that the product $f'_l \cdot f'_i \cdot f'_c$ can be simply written as the ratio $L_c / L_p$. The logic behind Eqs.(\ref{eq:fl})-(\ref{eq:fc}) is that in a steady state and a random observing time, the probability of seeing a short-duration event during a long-duration event should be the ratio between the durations of the former and the latter.

Plugging Eqs.(\ref{eq:fl})-(\ref{eq:fc}) into Eq.(\ref{eq:Drake2}), one can derive
\begin{eqnarray}
 N & = &  N_* \cdot f_p \cdot n_e^{\rm ceti} \cdot \frac{L_c}{L_p}  \nonumber \\
 & = & \frac{L_{\rm MW}}{L_p} \cdot \frac{N_*}{L_{\rm MW}} \cdot f_p \cdot n_e^{\rm ceti} \cdot L_c \nonumber \\
 & \simeq & \frac{L_{\rm MW}}{L_p} \cdot R_* \cdot f_p \cdot n_e^{\rm ceti} \cdot L_c \nonumber \\
& \sim & R_* \cdot f_p \cdot n_e^{\rm ceti} \cdot L_c  \label{eq:Drake3}
\end{eqnarray}
where $R_* \simeq N_* / L_{\rm MW}$ is the average star formation rate throughout the MW history (with the assumption that the number of dead stars is much smaller than the number of living stars, which is justified since low mass stars that contributed to the vast majority of the total number have not died yet, see \citealt{westby20} for detailed calculations), and $L_{\rm MW} \sim 13.5$ billion years is the current lifetime of the Milky Way galaxy. Since the average lifetime of earth-like planets $L_p$ is $\sim L_{\rm MW}$ (e.g. for Earth it is the lifetime of the Sun, i.e. $\sim 10$ Gyr), the last step in Eq.(\ref{eq:Drake3}) has a $\sim$ sign. One can see that last step roughly reproduces the Drake equation (\ref{eq:Drake}) noticing Eq.(\ref{eq:neceti}).

\section{A quantitative assessment of the CETI signal emission rate}\label{sec:quantitative}

The number $N$ of CETIs is not a direct measurable quantity. A human observer may be more interested in the signal detection rate of CETI signals (e.g. in units of \# per year all sky), which I define as $\dot N_{\rm s,o}$, where the subscripts `s' and `o' denote `signal' and `observer', respectively. Ultimately, one cares about the average signal emission rate per CETI, which I define as $\dot N_{\rm s,e}$, where the subscript `e' denotes `emitter'. The emitted signals of CETIs may not always be detected by humans on Earth. I therefore introduce a parameter $\xi_o$ to denote the average fraction of the CETI signals that are detectable by humans on Earth. One can then write 
\begin{eqnarray}
 \dot N_{\rm s,o} & = &  N \cdot (\xi_{o} \dot N_{\rm s,e})  \nonumber \\
 & = & N_* \cdot f_p \cdot n_e^{\rm ceti} \cdot \frac{L_c}{L_p}  \cdot (\xi_{o} \dot N_{\rm s,e}).
 \label{eq:new}
\end{eqnarray}
where the first line of Eq.(\ref{eq:Drake3}) has been used.  This equation is more helpful than Eq.(\ref{eq:Drake3}) for a quantitative assessment of CETI signals. We now break down the terms introduced in Eq.(\ref{eq:new}) and discuss how they may be constrained astronomically. The discussions on various parameters of the Drake equation and other modified forms can be also found in \cite{vakoch15} and many papers in the literature, e.g. \cite{burchell06,westby20} and references therein. 
\begin{itemize}
 \item $N_*$: the Sun's distance from the Galactic center ($d_\odot \sim 8.0\pm 0.5$ kpc) and its proper motion velocity ($v_\odot \sim 220 \ {\rm km/s}$ as measured from the dipole moment in the cosmic microwave background) allows one to estimate that the total mass within the solar orbit is $M_{\rm in} = d_\odot v_\odot^2 / G \sim 10^{11} M_\odot$. Deducting the contributions from dark matter, gas and dead remnants (e.g. black holes and neutron stars) that may not be helpful to harbor life, one can write the stellar mass within the solar orbit as $M_{\rm *,in} = f_* M_{\rm in}$, where $f_* < 1$ is the fraction of mass attributed to stars. Consider that the fraction of stars in the Milky Way that is within the solar orbit is $f_{\rm in} < 1$. The total stellar mass should be $M_* = M_{\rm *,in} / f_{\rm in}$. Since the initial stellar mass function has a steep slope $N(m_*) dm_* \propto m_*^{-2.3} dm_*$ for $m_* > 0.5 M_\odot$ and $N(m_*) dm_* \propto m_*^{-1.3} dm_*$ for $0.08 M_\odot < m_* < 0.5 M_\odot$ \citep[e.g.][]{kroupa01}, the number of stars is dominated by those stars with the minimum mass $m_{*,m} = f_m M_\odot$ with $f_m < 1$. The total number of stars can be finally estimated as 
 \begin{eqnarray}
 N_* & \sim & \frac{M_*}{m_{*,m}} =\frac{f_*} {f_m f_{\rm in}}  \frac{M_{\rm in}}{M_\odot} \nonumber \\
 & = & \frac{f_*} {f_m f_{\rm in}} 10^{11} \sim (10^{11} - 10^{12}).
 \end{eqnarray}
 A commonly quoted number is $N_*=2.5\times 10^{11}$ \citep[e.g.][]{westby20}.
 \item $f_p$: Here the meaningful fraction should be the fraction of stars that can have habitable planets (or habitable moons orbiting giant planets) in stable orbits long enough to develop life and CETIs. Modern exoplanet observations suggest that planets may be ubiquitous in stellar systems \citep[e.g.][]{howard12,burke15}. Since single stars are likely the ones to harbor stable planet orbits, we assign $f_p$ as the fraction of single stars, which ranges from 1/2 to 2/3 \citep{lada06}.  
 \item $n_e^{\rm ceti}$: This is the number with a large uncertainty. $n_e$ may be determined with precision when survey observations of planets from nearby stellar systems are carried out, e.g. with the Transiting Exoplanet Survey Satellite (TESS) mission \citep{ricker15}. Based on the Kepler \citep{borucki10} observations, the fraction of GK dwarfs with rocky planets in habitable zones may be around 0.1 \citep{burke15}. This may be considered as the upper limit of $n_e^{\rm ceti}$ when considering other factors (planet mass, metallicity, existence of a magnetosphere, existence of a Jupiter-like large planet to deflect comets, etc., see, e.g. \cite{lineweaver01}) that might be relevant for producing CETIs. Since it is far from clear what physical conditions are essential for the emergence of CETIs, this number should allow for a large uncertainty, from 0.1 all the way to very small numbers. We normalize this number to $n_e^{\rm ceti} \sim 10^{-3}$ in the following discussion, keeping in mind the large uncertainty involved.
 \item $L_p$: Humans, the only intelligent life we know in the universe, emerge 4.54 billion years after the formation of the planet. One may conservatively assume that 4.5-billion-year is the minimum timescale to develop CETIs. As a result, the host stars of Earth-like planets to harbor SETIs should have long lives, e.g. of the solar or later types (GK dwarfs or M dwarfs). The planets where CETIs are harbored should also have survived for a comparable lifetime as their parent stars \citep[see also][]{westby20}. As a result, $L_p$ is likely at least several billion years, as is the case of Earth. 
 \item $L_c$: This is a parameter with the largest uncertainty, and cannot be estimated using available astrophysical observations. Since humans have developed communicating technology for more than a century on Earth, $L_c$ should be at least $\sim 100$ yr, so that $L_c / L_p$ is at least $10^{-8}$. However, it has been widely speculated that CETIs can survive much longer than this time scale (if they do not destroy themselves) \citep[e.g.][]{burchell06,vakoch15}. Lacking any guidance, in the following we normalize the ratio $L_c/L_p$ (which is relevant in the problem) to $ \sim 10^{-4}$, which corresponds to $L_c \sim (10^5-10^6)$ yr for $L_p \sim (10^9-10^{10})$ yr. Notice that in principle, CETIs can ``re-appear'' after self-destroy. What matters in our problem is the total duration of the existence of CETIs. The duration of each CETI and the number of generations of CETIs in a planet is not relevant. As a result, $L_c$ defined here can be considered as the average total duration of CETIs on each planet in the Galaxy.
 \item $\dot N_{\rm s,e}$: This is the average signal emission rate per CETI (for detailed discussion of CETI signals from the emitter's perspective, see Sect. \ref{sec:signals}).  The CETIs may repeat their signals multiple times, and $\dot N_{s,e}$ is defined as the total amount of signals emitted by a CETI divided by its entire lifetime $L_c$, averaged over all CETIs.  Maybe (and likely) different CETIs attempt to communicate using different types of signals. Maybe the same CETI attempts to communicate using several different types of signals. So, $\dot N_{s,e}$ is signal-type-dependent, and should be defined for each type of signal specifically.
 \item $\xi_o$: In order to connect the emission rate with the detection rate, one should introduce this  factor that denotes the fraction of emitted signals detectable by astronomers on Earth. One may write
 \begin{equation}
     \xi_o \equiv \frac{\Delta\Omega_e}{4\pi} \left(\frac{d_{\rm lim}}{d_{\rm MW}}\right)^2,
     \label{eq:xio}
 \end{equation}
 which includes the average beaming factor $\Delta\Omega_e / 4\pi$ for the emitted signals and the flux limitation factor\footnote{A simple isotropic distribution is assumed here for order-of-magnitude estimation. A more careful study should account for the MW structure and the anisotropic environment of the Earth neighborhood.} $(d_{\rm lim} / d_{\rm MW})^2$, where $d_{\rm lim}$ is the maximum distance from Earth the signal can be detected and $d_{\rm MW} \sim 10$ kpc is the characteristic distance scale of the Milky Way galaxy. This second factor depends on the strength (luminosity) of the emitted signal and the sensitivity of the telescopes that detect these signals. Ideally, advanced CETIs may emit signals detectable by all other civilizations across the Galaxy. For such signals, one takes $(d_{\rm lim} / d_{\rm MW})^2 \sim 1$, so that $\xi_o \simeq \Delta\Omega_e/4\pi$, which only  depends on the average solid angle $\Delta\Omega_e$ of the emitted signals.
 \item $\dot N_{\rm s,o}$: This is the total detectable signal rate at Earth from all sky all time. Similar to $\dot N_{\rm s,e}$, it should be defined specifically for each type of signal. It is not the rate of the truly detected signals, which depends on the fraction of sky coverage and the duty cycle of the telescopes. With dedicated surveys with certain sky and temporal coverage, the detected signal rate (or, very likely, its upper limit) can be corrected to derive $\dot N_{\rm s,o}$ (or, very likely, its upper limit).
 \item $N$: Even though this is not a direct measurable quantity, it is nonetheless interesting to write down the estimated CETI number $N$ in the Galaxy according to Eq.(\ref{eq:Drake3}) with the normalization values of each parameter as discussed above:
 \begin{eqnarray}
     N & = & 12500 \left(\frac{N_*}{2.5\times 10^{11}}\right) \left(\frac{f_p}{0.5}\right) \left(\frac{n_e^{\rm ceti}}{10^{-3}}\right) \nonumber \\
     &\times& \left(\frac{L_c/L_p}{10^{-4}}\right).
 \end{eqnarray}
\end{itemize} 
Notice that if one chooses $L_c \sim 100$ yr, $L_p \sim 10^{10}$ yr, and $f_p \cdot n_e^{\rm ceti} \sim 10^{-2}$ (corresponding to the factor $f_{\rm L} \cdot f_{\rm HZ} \cdot f_{\rm M}$ defined by \citealt{westby20}), one obtains a minimum CETI number of 12.5, which is consistent with the minimum CETI number estimated by \cite{westby20}.

\section{CETI Signals from the emitter's perspective}\label{sec:signals}

Great efforts have been made in the Search-for-Extra-Terrestrial-Intelligence (SETI) community to speculate the types of the CETI signals. \cite{wright18} defined an eight-dimensional model to describe CETI signals and argued that the current SETI searches only touched a tiny phase space of this eight-dimention ``cosmic haystack''. Some of the dimensions defined by \cite{wright18} (e.g. their dimension 1 [sensitivity to transmitted or received power] and their dimensions 3-5 [distance and position]) are from the observer's perspective.

In the following, I discuss a possible CETI signal from {\em the emitter's perspective} by speculating what type of signal a CETI may emit. Assuming that CETIs communicate using an electromagnetic signal\footnote{In principle, a multi-messenger channel may be also used. However, these messengers are technically challenging and not economical. For example, the generation of gravitational waves is very expensive. Cosmic rays tend to be deflected by interstellar magnetic fields. Even if neutrinos may be easier to generate than the other two messengers, they are very difficult to detect from the observer's prospective due to the small cross section of weak interaction. These channels may not be favored by CETIs and are, therefore, not discussed in this paper.}, I characterize a CETI signal by the following seven parameters:
\begin{enumerate}
    \item Duration: Shorter durations may be preferred from economical considerations, but the signal should be long enough for other civilizations to detect.
    \item Peak luminosity: Since any signal has a rising and fading phase, the luminosity at the peak time is a relevant parameter to consider. This parameter will define how far the signal can be detected by other civilizations given a certain detector sensitivity. 
    \item Emission spectrum: This concerns the central frequency and the bandwidth of the transmission signal, which has been discussed by \cite{wright18} as their dimensions 2 \& 6.
    \item Polarization: The polarization properties of a CETI signal may carry additional information. This is the dimension \#8 of \cite{wright18}.
    \item Lightcurve: This includes how the luminosity of an individual signal rises and falls as well as how multiple signals group together to display  intelligent information. This is somewhat discussed as the dimension \#9 by \cite{wright18} as ``modulation''.
    \item Solid angle: This describes how wide the CETI signal beam is, which is directly related to $\xi_o$ discussed in Eq.(\ref{eq:xio}).
    \item Repetition rate: This defines the $\dot N_{s,e}$ in Eq.(\ref{eq:new}), which is the average value of all CETIs in the Galaxy. The dimension \#7 of \cite{wright18} concerns the detected repetition rate, which is related to $\dot N_{s,o}$ in Eq.(\ref{eq:new}).
\end{enumerate}

Since no CETI signal has been detected, one can only take humans' own communicating signals for comparison. The most famous signal  was the ``Arecibo message'' broadcast in 1974 \citep{Arecibo}. 
The properties of the signal in connection with the seven dimensions discussed above are:
\begin{itemize}
    \item Duration: $\lesssim$ 3 minutes;
    \item Luminosity: 450 kilo-Watts or $4.5 \times 10^{12} \ {\rm erg \ s^{-1}}$;
    \item Spectrum: At frequency 2,380 MHz with an effective bandwidth of 10 Hz, with modulations by shifting the frequency by 10 Hz;
    \item Polarization: No information is available.
    \item Lightcurve: Consisting of 1,679 binary digits (approximately 210 bytes) that encode rich information such as numbers, atomic numbers, DNAs, a human image, the solar system, and the Arecibo radio telescope.
    \item Solid angle: A narrow beam of about 1 square arc-minute pointing toward the globular cluster M13 $\sim 25,000$ light years away. 
    \item Repetition rate: Not repeated with the same configuration according to the public record.
\end{itemize}

One may estimate the detectability of this signal by other CETIs. First, aliens in all other directions outside the narrow beam would never know that such a signal was emitted. Second, at the distance of M13, the flux density of this signal is $\sim 2.7\times 10^{-20}$ Jy, which is 21 orders magnitude fainter than the sensitivity (10 Jy) of the current humans' own narrow-band SETI search instrument, the Allen Telescope Array \citep{AllenTelescope}. Suppose that there are indeed aliens $\sim$ 25,000 years later on a planet orbiting one of the stars in M13, they need to use a telescope with a collecting area much greater than the size of the planet itself in order to detect the signal. The telescope should be also pointing toward the direction of Earth during the $\lesssim$ 3 minutes of time when the radio wave passes by the planet. The chance is negligibly small. Finally, since no other star is along the path towards M13, the only civilizations other than those in M13 who could in principle receive the signal would be the even more distant ones behind M13. The flux on their planets are even much lower and even more advanced technology is required to catch the signal.

So, if CETIs do exist and indeed have the intention to broadcast their existence in the Galaxy, they need to emit signals that are many orders of magnitude more powerful and in a much larger solid angle than the Arecibo message signal humans sent.

\section{FRB-like artificial signals by CETIs}

Fast radio bursts (FRBs) are frequently detected, millisecond-duration, radio bursts that originate from cosmological distances \citep{lorimer07,petroff19,cordes19}. Since their physical origin is unknown, an alien connection was speculated by some authors. For example, when placing physical constraints on FRBs, \cite{luan14} discussed the possibility that the observed FRBs are artificial signals sent by aliens and concluded that a modest power requirement is needed. \cite{lingam17} interpreted FRBs as radio beam signals produced by extra-galactic advanced civilizations to launch light sails for interstellar travels. These suggestions intended to interpret at least a fraction of the observed FRBs as of an artificial origin.

I believe that the observed cosmological FRBs are due to  astrophysical origins. Indeed, many theoretical models have been proposed in the literature to interpret various observational properties of FRBs \citep[e.g.][]{platts19}. Rather, I speculate that if CETIs do exist and indeed have the intention to communicate, they may send off FRB-like artificial signals to the Galaxy, likely along the Galactic plane. There are two main reasons for this speculation. First, intelligent civilizations know that FRBs happen very frequently (of the order $\sim 10^4$ per day all sky for bright ones) in the universe and that other civilizations must be monitoring these events all the time. The same observational facilities designed to detect FRBs (e.g. wide-field and sensitive radio telescope arrays) could easily spot artificial signals they would send. Second, probably more importantly, FRBs have radio frequencies and extremely short durations (but still long enough for detection). They are relatively easy to mimic from the economical point of view. In contrast, it is for example more difficult to make artificial gamma-ray bursts or fast optical bursts. 

In terms of the seven dimensions discussed in Section \ref{sec:signals}, the properties of an FRB-like CETI signal may be speculated as follows:
\begin{itemize}
    \item Duration: An FRB-like signal should have a duration of the order of milliseconds, i.e. $\Delta t = (1 \ {\rm ms}) \ \Delta t_{\rm ms}$.
    \item Luminosity and energetics: How bright the signals are emitted depends on the technological level of the aliens, but an advanced CETI who is eager to broadcast its existence may try to emit at a luminosity such that the detected flux level by a typical Milky Way observer (like humans on Earth) is comparable to that of a cosmological FRB. For a 1 ms-Jy signal with a characteristic distance in the Milky Way, the isotropic luminosity should be $\sim  (10^{32} \ {\rm erg \ s^{-1}}) \ d_{\rm 10 kpc}^2$. Consider the beaming factor (which is also the probability factor for the observer to see the signal), $\xi_o = \Delta \Omega/4\pi$. The true emission power\footnote{This is much greater than the one estimated by \cite{luan14}.} should be $\sim (10^{29} \ {\rm erg \ s^{-1}}) \xi_{o,-3} d_{\rm 10 kpc}^2$ or $(10^{22} \ {\rm W}) \xi_{o,-3} d_{\rm 10 kpc}^2$, where $\xi_o$ is normalized to $10^{-3}$. The true energy is $\sim (10^{26} \ {\rm erg}) \xi_{o,-3}\Delta t_{\rm ms} d_{\rm 10 kpc}^2$ or $\sim (10^{19} {\rm J}) \xi_{o,-3} \Delta t_{\rm ms} d_{\rm 10 kpc}^2$. This energy is $\sim 10^{11}$ times of the emitted energy in the Arecibo message, which is comparable to the rest mass energy of $\sim 100$ kg matter. Emitting such a signal is beyond the current technological capability of humans. On the other hand, there is no fundamental physical barrier to prevent this from happening. 
    \item Spectrum and dispersion measure (DM): FRBs have been detected from $\sim$400 MHz to 8 GHz \citep{petroff19}. The observed FRB flux typically show a steep drop at high frequencies. The low frequency range also suffers from a few suppression effects such as plasma scattering and absorption, which are particularly severe in the Galactic plane. One natural frequency CETIs may consider to broadcast artificial FRB signals would be around the hydrogen 21cm line frequency (1.42 GHz). Its ``resonances'' (e.g. 1.5 times or 2 times) may be also possible. In order to mimic an FRB signal, the CETIs may consider to emit the signal in a wide enough bandwidth so that a DM could be measured. In order to draw attention to observers, they may also add an extra DM (e.g. $\sim 500 \ {\rm pc \ cm^{-3}}$, a typical value for cosmological FRBs) in excess of the Galactic value by placing a cold plasma along the path of signal propagation. Such a signal can be easily picked up by  observers using the standard FRB-searching algorithms. 
    \item Polarization: So far, the FRB polarization properties do not show well-defined common characteristics. The polarization properties are also not essential for FRB detections. There is no need to design certain polarization properties in the artificial signals unless some extra intelligent information can be carried. For example, a CETI may decide to emit an FRB with 100\% linear polarization with a Faraday rotation measure (RM) greatly exceeding the local value in the Galaxy through generating a strong magnetic field in the emission site. They may also vary RM significantly to show that the signals are indeed artificial.
    \item Lightcurve: This is how CETIs deliver information to show the intelligent nature of their signals. There are many possibilities, which we refrain from speculating. In any case, one may expect a series of FRB-like signals that encode profound information understandable by other advanced civilizations. 
    \item Solid angle: Advanced CETIs may like to broadcast their signals as wide as possible, so that an all-sky ($4\pi$) signal would be most ideal. In practice, it would be easier to send collimated signals for economical and technical reasons. Since the Galactic plane has the highest probability for other civilizations to detect the signals, a CETI in the Galactic plane may send a fan beam with the vertical angle defined by the height to radius ratio of the Galactic plane, which is of the order of $10^{-3}$. If the azimuthal angle is $2\pi$, the beaming factor is $\Delta\Omega/4\pi \sim 10^{-3}$.
    \item Repetition rate $\dot N_{s,e}$: This is something not easy to estimate, but can be in principle constrained from the data (see Section \ref{sec:quantitative-FRB} below for detailed discussion). However, since sending these signals consumes a lot of energy, the emission rate may not be very frequent unless CETIs are in great desire to communicate (e.g. sending S.O.S signals for help).
\end{itemize}

\section{A Quantitative Assessment of  CETI's FRB-like Artificial Signals}\label{sec:quantitative-FRB}

With the operations (or planned operations) of a growing number of radio antenna arrays (e.g. The Canadian Hydrogen Intensity Mapping Experiment [CHIME] \citep{CHIME}, the Deep Synoptic Array 2000-antenna [DSA-2000] \citep{DSA-2000}, and the Square Kilometre Array [SKA] \citep{SKA}) to detect FRBs, one can start to place a constraint on $\dot N_{\rm s,o}$ of FRB-like artificial signals from the Milky Way. Non-detection of any FRB-like artificial signal from the Galactic plane, when corrected for the sky coverage and duty cycle for progressively longer observational times, can place progressively tighter constraints on $\dot N_{\rm s,o}$, which would place constraints on the average signal emission rate of CETIs, $\dot N_{\rm s,e}$, for FRB-like signals based on Eq.(\ref{eq:new}). For example, an upper limit of $\dot N_{\rm s,o} < 0.1 \ {\rm yr}^{-1}$ (e.g. all sky no detection in a decade) can lead to a  constraint
\begin{eqnarray}
    \dot N_{\rm s,e} & < & (0.008 \ {\rm yr^{-1}}) \left(\frac{\dot N_{\rm s,o}}{<0.1 \ {\rm yr^{-1}}}\right) \left(\frac{N_*}{2.5\times10^{11}}\right)^{-1}  \nonumber \\
    &  &\left(\frac{f_p}{1/2}\right)^{-1} \left(\frac{n_e^{\rm ceti}}{10^{-3}}\right)^{-1} \left(\frac{L_p/L_c}{10^{4}} \right) 
    \left(\frac{\xi_o}{10^{-3}}\right)^{-1}, \nonumber \\
\label{eq:new2}
\end{eqnarray}
where the parameters are normalized to the characteristic values as discussed in Section \ref{sec:quantitative}. With the non-detection upper limit in decades, one may then make a quantitative statement that with the fiducial values of the parameters, CETIs on average emit less than one per century FRB-like artificial signals that can cross the entire Milky Way. 

\section{Fermi-Hart paradox}

One commonly discussed question in the SETI community is the  ``Fermi-Hart paradox'' regarding ``where is everybody?'' (as elaborated in detail by \cite{hart75}, see also \cite{brin83}). Even though the question was about why there is no evidence for extraterrestrial intelligence visiting Earth, another version was to address why we have not received signals from CETIs. One general type of answer to this question is that humans have not observed the universe long enough to allow any detection. Indeed, \cite{wright18} showed that humans only searched a tiny parameter space in a multi-dimensional ``cosmic haystack'' through blind searches. 

If one focuses on one specific type of signal, e.g. the FRB-like artificial signal discussed in this paper, the ``haystack'' search volume greatly shrinks. I argue that even for such specific signals, the answer to the ``Fermi-Hart paradox'' is still {\em ``we simply have not observed long enough yet''}. One can quantitatively show this using Equation (\ref{eq:new2}). For FRB-like signals, even if one can achieve the $\dot N_{\rm s,o} < 0.1 \ {\rm yr}^{-1}$ upper limit (which requires dedicated efforts from wide-field radio arrays working for decades), for fiducial parameters, one can only set a moderate upper limit on the signal emission rate of highly advanced CETIs who can broadcast their existence across the entire Milky Way, i.e. $\dot N_{\rm s,e} < 0.008 \ {\rm yr}^{-1}$. It is impossible that all CETIs have such a capability to broadcase across the Milky Way (e.g. humans do not). If CETIs are less advanced, $\xi_o$ would be greatly reduced since the $(d_{\rm lim}/d_{\rm MW})^2$ factor would have to be included. For example, for $\xi_o \sim 10^{-7}$ (corresponding to the case that detectable CETI signals can only reach a distance of $d_{\rm lim} \sim 100$ pc), one can only set a limit of $\dot N_{\rm s,e} < 80 \ {\rm yr}^{-1}$ for $\dot N_{\rm s,o} < 0.1 \ {\rm yr}^{-1}$ and typical parameters. It is hard to imagine that an average CETI so desperately communicates with the universe by sending signals at a rate $>80 \ {\rm yr}^{-1}$. As a result, there is essentially no ``paradox''.

This argument applies not only to the FRB-like signal but also other specific signals, which are probably even more difficult to generate by CETIs. If one does not specify a signal type, the search parameter volume would be increased exponentially, so that the ``paradox'' is further diminished \cite[][and references therein]{wright18}. The lack of detection of any CETI signals now is naturally expected.

\section{Summary and discussion}

The points made in this paper can be summarized as follows:

\begin{itemize}
    \item I presented a derivation of Equation (Eq.(\ref{eq:Drake3})) based on a probability argument, which is consistent with the original Drake equation (Eq.(\ref{eq:Drake})).
    \item Based on Eq.(\ref{eq:Drake3}), I proposed a new equation (Eq.(\ref{eq:new})) to connect the observed CETI signal rate $\dot N_{s,o}$ with the average signal emission rate $\dot N_{s,e}$ by CETIs. Subject to uncertainties of several parameters, this equation allows one to use observations to directly infer the value (or, very likely, the upper limit) of the average CETI signal emission rate. The equation applies to specific signals rather than unspecified blind-search signals. 
    \item After characterizing CETI signal properties in seven dimensions from the emitters' perspective, I suggested that FRB-like artificial signals could be one type of CETI signals for good reasons. Using Eq.(\ref{eq:new}), I derive a constraint one may pose with the detection/non-detection of FRB-like artificial signals from the Milky Galaxy. 
    \item The $\dot N_{s,e}$ constraint derived from the detection/non-detection of FRB-like artificial signals (Eq.(\ref{eq:new2})) is taken as an example to quantitatively show why the ``Fermi-Hart paradox'' is not a concern. Even for one particular type of signal and under the most optimistic assumption (i.e. an average CETI is able to broadcast FRB-like artificial signals across the entire Milky Way), one would not expect to detect any signal now and probably still not even after decades or centuries of dedicated monitoring. This strengthens the argument against the Fermi-Hart paradox by \cite{wright18} for blind searches.
\end{itemize}

Finally, one natural question is whether humans should send FRB-like artificial signals in the future when technology is advanced enough. This is definitely a topic subject to debate (see, e.g. \cite{Gertz16} for a general discussion on the pros and cons on Messaging to Extra-Terrestrial Intelligence (METI)). Optimists may think that CETIs are eager to find out whether they are alone in the universe and would be happy to remotely communicate with other civilizations. Pessimists, on the other hand, would believe that it is very dangerous to expose ourselves to more advanced civilizations as they would invade Earth to snatch resources\footnote{This point was raised by Stephen Hawking in his 2010 documentary series and delineated in Liu Cixin's famous science fiction trilogy ``Remembrance of Earth's Past'' (or ``The Three-Body Problem'' series). See also \cite{Gertz16} for a critical review.}. In any case, I believe that this should be a decision to be made by the entire humanity, not by a small group of ``elites''. My personal recommendation is: Do not do anything until one can develop the technology to emit FRB-like signals (this may take some time, e.g. hundreds, thousands or even millions of years); keep watching whether ``others'' have emitted any such signal along the way; and make a decision then! If non-detection of FRB-like signals persists for a long time (e.g. after thousands of years), then the Fermi-Hart paradox may become more a concern. The CETIs may be also taking a pessimistic approach like us, so that the paradox may find an answer along this reasoning in the far future.

\bigskip
I thank Qiang Yuan for an important remark on an earlier version of the paper, Jason Steffen for discussing the current status of exoplanet searches, and Maura McLaughlin for discussing the current status of Galactic FRB-like signal searches. 


\end{document}